# Damageless Tough Hydrogels with On-demand Self-reinforcement

Chang Liu[1], Takeshi Fujiyabu[2], Naoya Morimoto[1], Lan Jiang[1], Hideaki Yokoyama[1], Takamasa Sakai[2], Koichi Mayumi[1, 3]*, Kohzo Ito[1]*

[1] Material Innovation Research Center (MIRC) and Department of Advanced Materials Science, Graduate School of Frontier Sciences, The University of Tokyo, 5-1-5 Kashiwanoha, Kashiwa, Chiba 277-8561, Japan.

[2]Department of Bioengineering, Graduate School of Engineering, The University of Tokyo, 7-3-1 Hongo, Bunkyo-ku, Tokyo, Japan.

[3]AIST-UTokyo Advanced Operando-Measurement Technology Open Innovation Laboratory (OPERANDO-OIL), National Institute of Advanced Industrial Science and Technology (AIST), 5-1-5 Kashiwanoha, Kashiwa, Chiba 277-8561, Japan.

*Corresponding author. Email: kmayumi@molle.k.u-tokyo.ac.jp (K.M.); kohzo@edu.k.u-tokyo.ac.jp (K.I.)

Abstract

Most tough hydrogels are reinforced by introducing sacrificial structures that can dissipate input energy. However, since the sacrificial damages cannot recover instantly, the toughness of these gels drops substantially during consecutive cyclic loadings. Here, we propose a new damageless reinforcement strategy for hydrogels utilizing strain-induced crystallization (SIC). In Slide-Ring (SR) gels with freely movable cross-links, crystalline repetitively forms and destructs with elongation and relaxation, resulting in both excellent toughness of 5.5 ~ 25.2 $MJ/m^3$ and 87% ~ 95% instant recovery of extension energy between two consecutive 11-fold loading-unloading cycles. Moreover, SIC occurs "on-demandly" at the crack-tip area where strain amplification and stress concentration take place and forces the crack to turn sideways. The instantly reversible tough hydrogels are promising candidates for applications in artificial connective tissues such as tendon and ligament.

**Main Text**

Hydrogels are three-dimensionally cross-linked polymer networks preserving large amount of water. Because of their high water content and biocompatibility, hydrogels display promising prospect in biomedical applications. However, except for those with low requirement for mechanical strength such as contact lenses and liquid bandages, further applications, such as connective and supportive artificial tissues like tendon, ligament and cartilage, has been restricted due to the poor mechanical properties of common synthetic hydrogels.

To improve the mechanical toughness of hydrogels, people utilize the reinforcement strategy of integrating energy dissipation mechanisms, such as sacrificial networks (*1-5*), nanoparticles (*6, 7*), and highly ordered crystalline structures (*8-10*) into gel network. Under large deformations, the "subtractive" damage of these structures dissipates input strain energy and raises the apparent work required for rupturing the whole material. Moreover, by utilizing reversible physical interactions as the energy dissipation mechanism, reusable tough gels with high extent of structural and mechanical recovery can be realized (*2, 3, 7, 9*). A very recent study (*5*) even realizes more than 100% mechanical recovery during repetitive loading by initiating new polymerization from the mechanoradicals that generated by sacrificially damaged structures. However, the reconstruction process of the damaged structures usually takes minutes or hours. The more the energy dissipation is, the longer the reconstruction time will be. As a result of such "trade-off" relationship between toughness and instant recovery, substantial deterioration in mechanical strength is commonly observed when the tough gels undergo two consecutive loading-unloading cycles with no waiting time. This is undesirable for the applications like connective and supportive artificial tissues, which require robust mechanical behavior against cyclic large deformations.

### The "additive" concept

Instead of relying on the "subtractive" rupture of existing structures that requires long time for their reconstruction, we propose a damageless reinforcement strategy utilizing the "additive" concept, i.e., the formation of new structures under stretching. As shown in Fig. 1A and 1B, new structures composed of highly oriented polymer chains appear "on-demandly" at large deformations or at the areas that encountering stress concentration, e.g., the tip of an artificial crack. These strain-induced structures work as nano fillers and enhance the overall strength of the material. Moreover, since the structures are surrounded by water and are stable only under large deformations, their forming and destruction process are thought to be highly reversible and respond instantly to the loading condition. In this way, robust hydrogels that maintains good mechanical toughness against cyclic stretching can be realized.

In practice, we focused on two well-known polyethylene glycol (PEG) hydrogels invented and developed in the past decade, Tetra-PEG gel (Fig. 1C) and Slide-Ring (SR) gel (Fig. 1D). Tetra-PEG gel and SR gel are chosen based on two considerations: PEG chains have delicate inter-chain interaction that can lead to the formation of close-packing structure (*11*); moreover, stress can be homogenized across the network of both gels so that they can undergo large deformations to extend and orient the PEG chains. However, it is also worth to mention that the two kinds of gels have different principles of homogenizing the stress distributions. Tetra-PEG gel, fabricated from two mutually reactive four-armed pre-polymers, has well-defined and ideally homogeneous network at the as-prepared state (*12*). SR gel, on the other hand, contains movable figure-of-eight cross-links that work as pulleys to eliminate stress heterogeneities during the deformation (*13*). The figure-of-eight cross-links are topologically confined by the adjacent ring molecules threaded by the same PEG chains. Therefore, the smaller the amount of ring molecules in one PEG chain is, the wider the slidable range of cross-links is, and the higher the extents of PEG chain extension and interaction are (*14*). In this study, we intentionally decreased the number of ring molecules so that they only cover 2% of the PEG monomer units within the SR gels. The polymer concentration of Tetra-PEG gel is 16 wt%, whilst that of SR gel is adjusted to 25 wt% so that it has similar PEG-water fraction with Tetra-PEG gel. Detailed calculations are given in the supplementary materials.

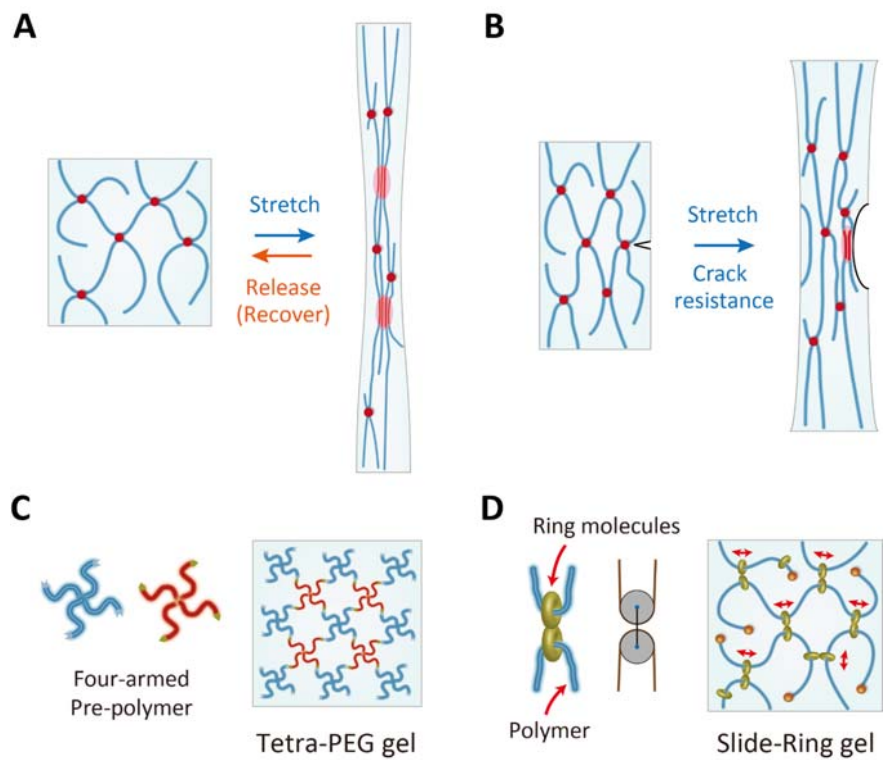

Fig. 1. Schematic illustrations of "additive" reinforcement strategy. (A) The formation and destruction of the strain-induced structure during the cyclic loading-unloading of a stripe-shaped hydrogel. (B) The formation of strain-induced structure at the crack-tip area of a single edge notched tension (SENT) hydrogel specimen. (C) Mutually reactive four-armed pre-polymers and the Tetra-PEG gel. (D) Figure-of-eight cross-links formed by two ring molecules on different polymer chains and the Slide-Ring (SR) gel.

Instantly reversible reinforcement

The loading-unloading behavior of both gels at their maximum extensions is shown in Fig. 2A. The stretching behavior of Tetra-PEG gel (16 wt%) is well predicted by the Neo-Hookean model (*15*) in the range of $1.0 < \lambda < 3.0$ owing to its highly homogeneous network. When $\lambda > 3.0$, stress exceeds the predicted value due to the limited chain extensibility. During the whole process, Tetra-PEG gel does not show any detectable hysteresis. In contrast, SR gel (25 wt%) shows hysteresis loop in the high $\lambda$ region ($5.0 < \lambda < 13.0$) of the loading-unloading curve, and the ratio of the area covered by the hysteresis loop to that covered by the loading curve is less than 5%. Moreover, as shown in Fig. 2B, such hysteresis loop occurs repeatedly during five consecutive loading cycles ($1.0 < \lambda < 11.0$) conducted in a custom-ordered liquid paraffin chamber to prevent the evaporation of the solvent (Fig. S1). This unique behavior suggests that the energy dissipative structural change in highly stretched SR gel is instantly reversible.

In Fig. 2C and movie S1, we show the crack growth behavior of both gels in pure shear (PS) geometry. Tetra-PEG gel ruptures almost simultaneously with crack initiation at $\lambda = 1.6$, and the crack propagates perpendicularly to the stretching direction (black arrow) at the velocity of 292 mm/s. In contrast, the crack propagation velocity in SR gel is 5 mm/s, which is only 1/60 of that of Tetra-PEG gel, and its rupture is delayed to $\lambda = 3.2$. Furthermore, crack turning (red arrows) and branching (blue arrow) are observed in the PS specimen of SR gel. In the single edge notched tension

(SENT) specimen that has less restraint to lateral direction contraction, the crack propagation direction even turns 90 degrees (Fig. 2D and 2E). Commonly, large strain amplification and stress concentration exist at the crack-tip (*16*) and drive the crack to grow forward. The crack turning behavior of SR gel suggests anisotropic toughness that the fracture resistance in the forward direction is much higher than that in sideway directions. This unique behavior suggests that the "strain-induced" reinforcement occurs in SR gel.

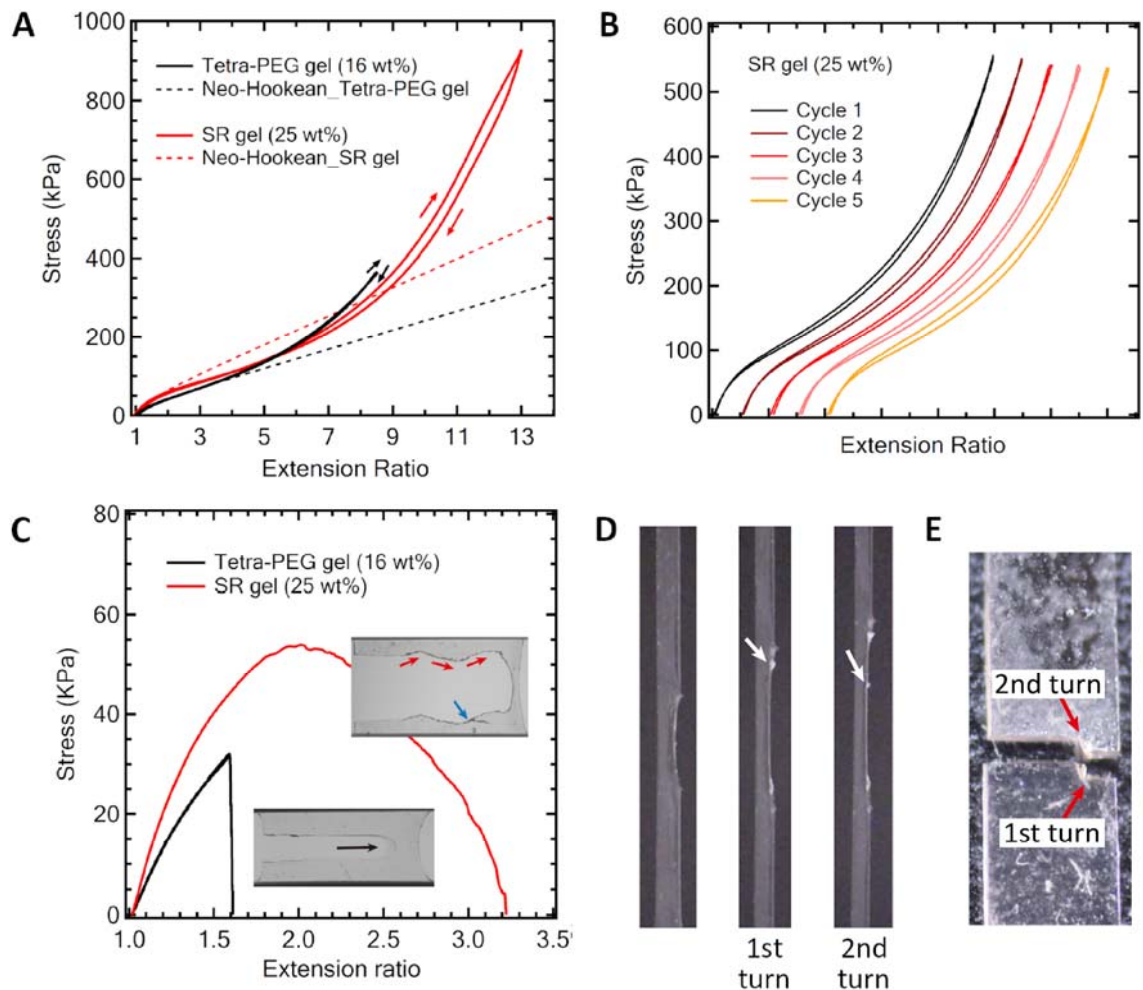

Fig. 2. Instantly reversible reinforcement under extreme extension. (A) Loading-unloading curves of Tetra-PEG gel and SR gel in the range 1.0 < $\lambda$ < 13.0 and their Neo-Hookean fitting results. (B) Hysteresis loops of five consecutive loading-unloading cycles of SR gel in the range 1.0 < $\lambda$ < 11.0. Horizontal shifts are applied to the curves of the 2nd, 3rd, 4th, and 5th cycles, respectively. (C) Stress-extension ratio curves of fracturing the pre-notched PS specimens of Tetra-PEG gel and SR gel. The insets are the photos of the propagating cracks; (D) Photos of crack turning during the fracture process of the pre-notched SENT specimen of SR gel. (E) Photo of the ruptured SENT specimen of SR gel.

### Instantly reversible structural transformation

In-situ wide-angle and small-angle X-ray scattering (WAXS and SAXS) are further conducted to understand the mechanism of the "strain-induced" reinforcement. Amorphous halo is observed in the 2D WAXS patterns of both gels at undeformed state (Fig. 3A and Fig. S2A, $\lambda$ = 1.0), and the halo turns into a broad peak at $q$ = 1.90 Å$^{-1}$ in the circular averaged 1D WAXS profiles shown in Fig. 3B and Fig. S2B. Such peak can be assigned to the hydrated PEG chains (*17*), and the decrease in the peak intensity corresponds to the reduction of hydration degree of PEG under elongation. The conformation of a stretched PEG chain in water changes gradually from 'random coil' to 'helix', and even to 'planar zig-zag' (*18*). During this process, the distance between the adjacent oxygen atoms in PEG increases and the interaction between PEG and water decreases.

Diffraction spots (Fig. 3A, blue arrows) appear in the 2D WAXS pattern of the extremely deformed SR gel, suggesting the formation of strain-induced crystalline. The spots turn into two peaks at $q$ = 1.45 Å$^{-1}$ and 1.73 Å$^{-1}$ in the 1D profile (Fig. 3B), respectively representing the $d$-spacing of 4.33 Å and 3.63 Å that can be indexed as (100) and (010) reflections of planar zig-zag PEG crystalline with triclinic unit cells (*19*) shown in Fig. 3C. This suggests that the PEG chains in the SIC of SR gel are fully extended and take all-*trans* conformation at $\lambda = 13.0$, instead of the *trans-gauche-trans* conformation in $-CH_2-CH_2-O-$ sequence that is usually observed in common monoclinic crystals (*20*).

Sharp streaks are observed in the 2D SAXS patterns of SR gel at $\lambda = 9.0$ and $\lambda = 13.0$ (Fig. 3A, yellow arrows). Such phenomenon is commonly observed in the shear-induced crystallization process of polymer melts such as polyethylene (*21*) and polypropylene (*22*), indicating the formation of crystallization precursor structure that consists of highly extended polymer chains. Our results for the first time show that similar level of chain extension can exist in SR hydrogel. By analyzing the streak profile at $\lambda = 13.0$ with Ruland's method (*23*) and Guinier law for long rod-like particles (*24*), the average length and radius of the SIC structure are estimated to be 100 nm and 8 nm, respectively (Fig. S3 and Fig. S4), which are both of the same order to those of previously reported strain-induced and shear-induced crystallines (*21, 22, 24, 25*).

It is worth to mention that, for Tetra-PEG gel, the X-ray scattering patterns obtained from the loading process is almost identical with those obtained at the same extension ratio of unloading process and no diffraction spots are observed (Fig. S2). However, slight hysteresis can be observed between the 1D profiles of SR gel obtained from loading and unloading process, which are respectively shown as solid and dash lines in Fig. 3B. Such hysteresis corresponds well to the 5% strain energy dissipation of SR gel. These results suggest that the hysteresis loop in the stress-extension ratio curves and the "strain-induced" reinforcement effect originate from the formation and destruction of SIC structures.

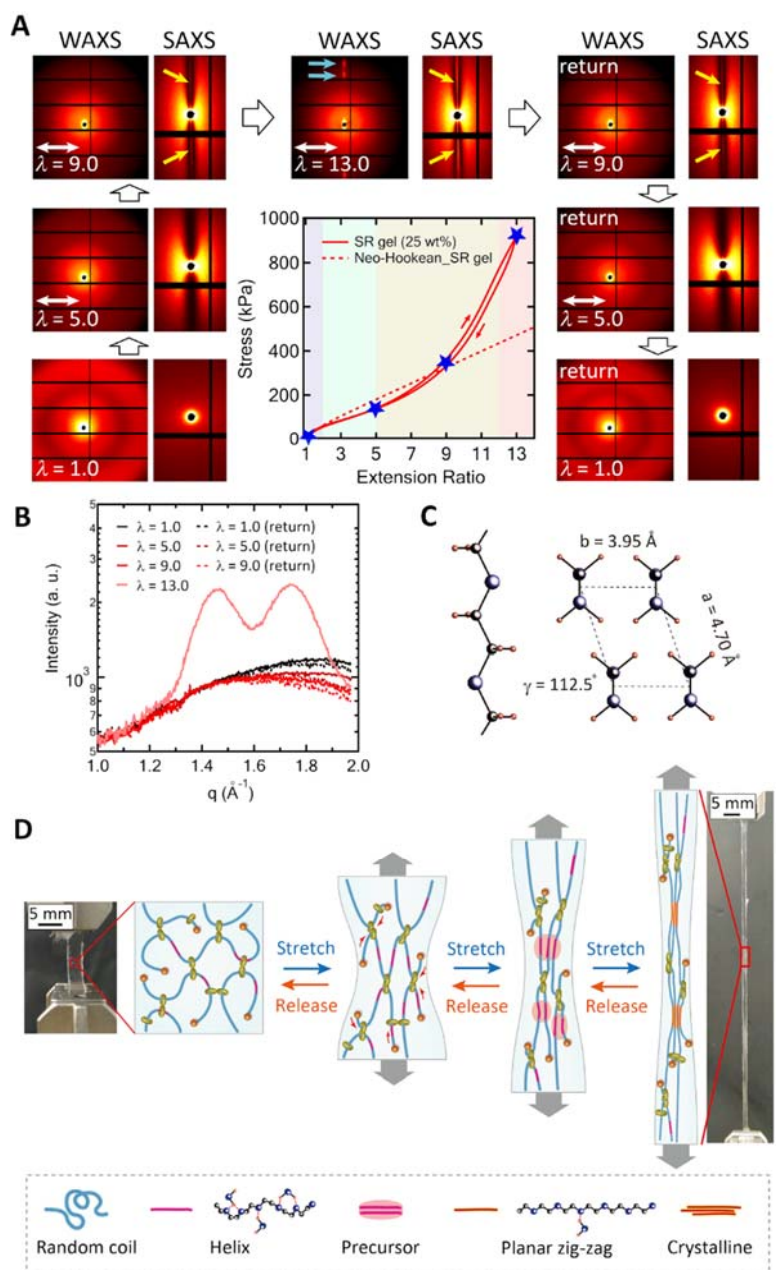

Fig. 3. Structural characterization of SR gel (25 wt%) under cyclic deformation. (A) WAXS and SAXS patterns obtained from the $1.0 < \lambda < 13.0$ loading-unloading cycle. The white double arrows denote the stretching direction, the blue and yellow arrows point out the diffraction spots in WAXS pattern and the sharp streaks in SAXS patterns, respectively. (B) WAXS 1D profiles obtained via 5° symmetric sector average in the perpendicular direction. (C) Structures of planar zigzag PEG and its triclinic crystal. (D) Schematic illustrations of the four-staged structural transformation.

Based on the abovementioned results, the structural transformation of a SR gel under uniaxial loading-unloading cycle is speculated as follows (Fig. 3D): (i) At $0 < \lambda < 2.0$, most of the PEG chains are in the random coil conformation, the elasticity of SR gel is governed by the entropy of PEG chains that affinely deforms with global elongation; (ii) At $2.0 < \lambda < 5.0$, the sliding movement of cross-links enlarges the apparent length of network strand, and causes the strain-softening of SR gel; (iii) At $5.0 < \lambda < 12.0$, the conformation of PEG chains deviates from Gaussian approximation and the enthalpic contribution to network elasticity starts to take place. Combining the effect of the formation of the rod-like crystalline precursors at $\lambda > 9.0$, strain-hardening is observed; (iv) At $\lambda > 12.0$, PEG chains are further extended to planar zig-zag conformation and finally form SIC structures. The formation of crystalline structures leads to the stress relaxation of non-crystalline network strands (*26*) and thus another slight strain-softening appears in the stress-extension ratio curve. (v) The reverse processes, such as the melting of crystalline, the destruction of crystalline precursors, and the returning of movable cross-links, take place during the unloading. The ring molecules can move freely so that almost no energy is dissipated when $0 < \lambda < 5.0$. However, the formation and destruction of precursor and crystalline structures involve PEG-PEG and PEG-water interactions, thus small amount of strain energy dissipation is observed when $5.0 < \lambda < 13.0$.

The unique SIC behavior of SR gel emphasizes the critical role played by the movable cross-links. Indeed, Tetra-PEG gel has well-designed network structure to reduce the stress concentration, and all the network strands can be fully stretched at the same time to tolerate extreme deformations. However, for Tetra-PEG gel with fixed cross-links, stress concentration due to meso- or macro-scale cracks is inevitable in practical situations. Catastrophic rupture may occur at these places before the chains approach each other to form crystalline. In contrast, SR gel can automatically adjust its network structure corresponding to the local stress distribution. The strongest stress concentration not only results in the highest extent of chain alignment, but also leads to the most severe sliding movement of cross-links, which increases the number of monomers in PEG strands between two cross-links. These structural changes ensure the efficient PEG-PEG interaction and promote the SIC behavior.

### Effect of polymer concentration

As shown in Figs. 4A and 4B, SR gels with 40 and 50 wt% polymer concentration exhibit not only higher Young's modulus and work of rupture, but also stronger energy dissipation than SR gel (25 wt%). Since all the gels have the same cross-linker concentration, the increase of stiffness, toughness, and hysteresis can be attributed to the enhanced PEG-PEG interaction by increasing polymer concentration. We also notice in Fig. 4B that the hysteresis loop obtained from the second cycle is smaller than that obtained from the first one. That is, part of the structural change during stretching is not instantly reversible and dissipates energy only in the first loading cycle. Such phenomenon is usually reported as the "Mullins effect" in filler-reinforced rubbers and unfilled ones that crystallize (*27*). Nevertheless, the hysteresis loops from the second to the fifth cycle overlap with each other, and the amount of extension energy stored during each loading process is maintained at 87% of that stored during the first stretching. The value is still larger than that of conventional gels. This part of reversible structural change, which consistently dissipates energy from the first to the fifth cycle, still plays a critical role in preserving the mechanical robustness of the SR gel.

The effect of polymer concentration on the structural transformation behavior of SR gels is also reflected by the WAXS results shown in Figs. 4C and 4D. Diffraction spots appear at $\lambda = 7.0$ for SR gel (50 wt%), earlier than the case of SR gel (25 wt%). Moreover, the diffraction spot in the perpendicular direction turns into a peak at $q = 1.36$ Å$^{-1}$ in 1D profile, which can be assigned to the (120) reflection of the monoclinic crystal consisted of helical PEG chains (*28*) instead of fully stretched planar zig-zag ones. That is, enhanced PEG-PEG interaction accelerates the SIC formation. On the other hand, these diffraction spots still exist at $\lambda = 5.5$ during retraction, suggesting that the enhanced PEG-PEG interaction delays the SIC destruction as well. Regardless of the hysteresis within one loading-unloading cycle, it is worth to emphasize that the SIC of SR gel (50 wt%) maintains high reversibility among multiple cycles, as indicated by the identical WAXS profiles at the same stretching state of each cycle.

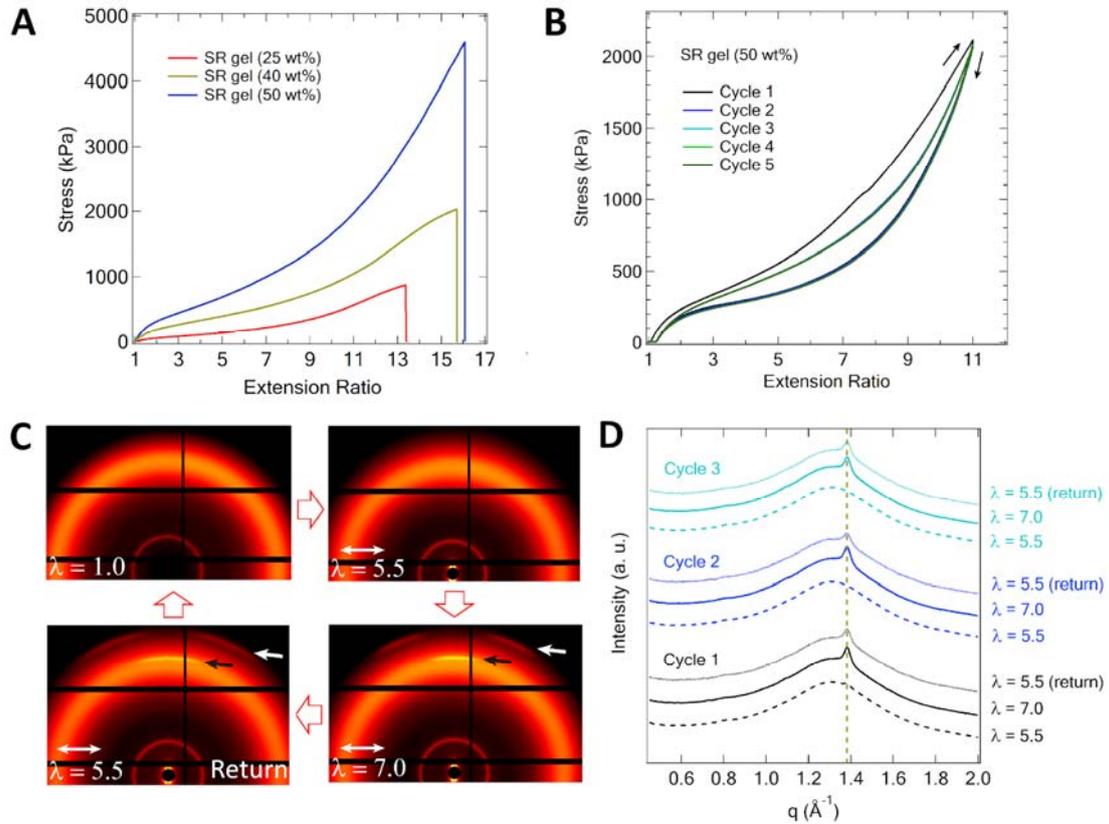

Fig. 4. Effect of polymer concentration on the toughness and the instant reversibility of SR gels. (A) Stress-extension ratio curves of SR gels with various polymer concentrations. (B) Stress-extension ratio curves of SR gel (50 wt%) obtained from five consecutive $1.0 < \lambda < 11.0$ loading-unloading cycles. (C) 2D and (D) 1D WAXS results of SR gel (50 wt%) obtained from $1.0 < \lambda < 7.0$ loading-unloading cycles. The black and white arrows denote the diffraction spots of monoclinic crystal of helical PEG.

Work of rupture vs. instant reversibility

In Fig. 5, we present a scatter plot of mechanical toughness against robustness of Tetra-PEG gel, SR gels, and some existing tough gels (*1, 3, 9, 10, 29-41*). The mechanical toughness is represented by the work of rupture deduced from the uniaxial tensile test. The instant reversibility, on the other hand, is defined as the extension energy ratio between two consecutive loadings. For some of the references (*1, 3, 30, 36, 39, 40*) that only provide the data of one loading-unloading cycle, the extension energy ratio of the unloading process to the loading process is used.

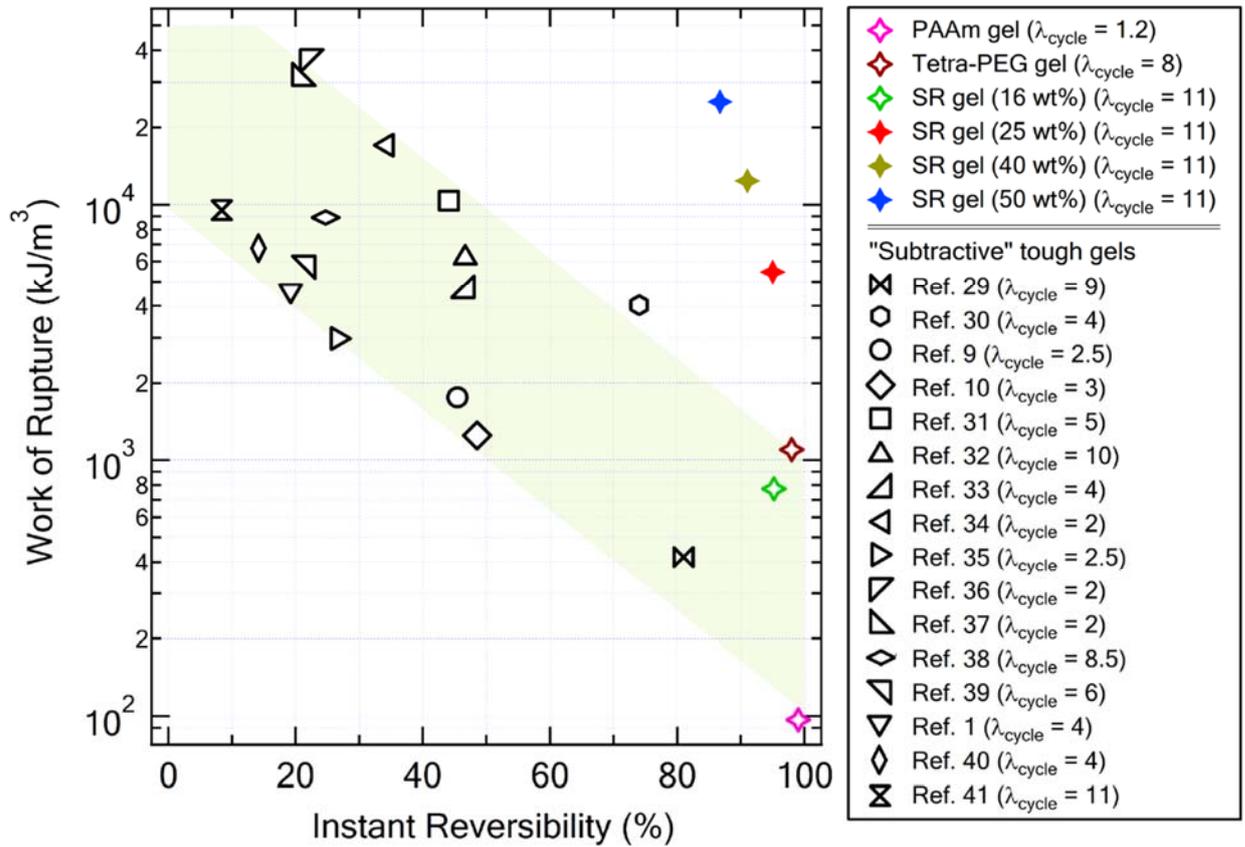

**Fig. 5. Work of rupture vs. instant reversibility plot of various kinds of gels compared with those of Tetra-PEG and SR gels.** The shaded area indicates the "trade-off" region.

As a representative of conventional chemical gels, polyacrylamide (PAAm) gel (*3*) shows almost complete reversibility but low mechanical toughness. The homogeneous network structure enhances the toughness of Tetra-PEG gel (16 wt%) and SR gel (16 wt%) (*14*) to some extent, but their data points still appear at the lower right corner of Fig. 5. On the other hand, the data points of existing tough gels that have superior toughness but poor reversibility, such as the double-network (DN) gels (*41*), appear at the upper left side.

By widening the movable range of cross-links and increasing the concentration of network polymer, SR hydrogels reinforced with instantly reversible SIC structures are for the first time realized. Wide slidable range of cross-links ensured the gel to endure large deformations, during which PEG chains are gradually stretched and aligned. High polymer fraction enhances the interaction of PEG chains that are exposed by the sliding of cross-links and further promoted the formation of oriented close packing structures in PEG-rich domains. With the help of these "on-demand" network structure transformation, the data points of SR gels appear at the upper right side of Fig. 5, deviating from the shaded "trade-off" region. This "additive" reinforcement concept will also be applicable to various water-soluble polymers that tend to crystallize, such as polyvinyl alcohol (PVA), and provides new insights into the practical applications of tough hydrogels.

**Acknowledgments:** We thank Dr. Taiki Hoshino, Dr. So Fujinami, and Dr. Tomotaka Nakatani (RIKEN SPring-8 center) for the valuable discussions and their generous effort in beamline setup. **Funding:** This work was partially supported by the ImPACT Program of the Council for Science, Technology, and Innovation (Cabinet Office, Government of Japan), a Grant-in-Aid for Young Scientists (B) (No. 15K17905), JST-Mirai Program Grant Number JPMJMI18A2, AIST-UTokyo Advanced Operando-Measurement Technology Open Innovation Laboratory (OPERANDO-OIL), and JSPS KAKENHI Grant Number JP 18J13038. **Author contributions:** C.L., T.S., K.M., and K.I. conceived the concept and designed the experiments. C.L., T.F., and N.M. performed the experiments. T.F., L.J., and H.Y. assisted with the material fabrication and characterization. C.L., K.M., and K.I. wrote the manuscript. All authors discussed the results and commented on the manuscript. **Competing interests:** Authors declare no competing interests. **Data and materials availability:** All data is available in the main text or the supplementary materials.

Supplementary Materials

Materials and Methods

Preparation of Tetra-PEG and SR gels

Tetra-PEG hydrogel was prepared by mixing two kinds of mutually reactive four-armed PEG solutions with 16 wt% concentration (Tetraamine-terminated poly(ethylene glycol) and Tetra-N-hydroxysuccinimide-glutarate-terminated poly(ethylene glycol), $M_w$ ~ 20000 g/mol, 5000 g/mol per arm, NOF Corporation, Japan). SR hydrogel was prepared from 2 % coverage polyrotaxane (PR-02, $M_w$ ~ 53000 g/mol), in which only 2 % of its PEG axis ($M_w$ ~ 30000 g/mol) was threaded with α-cyclodextrin (α-CD) rings. PR-02 was dissolved in deionized water and cross-linked by 3 wt% of divinyl sulfone (DVS, Tokyo Chemical Industry Corporation, Japan). The polymer concentration of PR-02 solution was adjusted to be 25, 40, and 50 wt%. Both gelation reactions took place in a 1 mm thick mold at room temperature for 8 hours.

PEG-water fraction in Tetra-PEG and SR gels

Since the molecular weight ratio between PEG axis and α-CD rings in PR-02 is 3/2.3, the weight concentration of PEG in SR gel (25 wt%) is 14.2 wt%. Therefore, the PEG-water fraction of SR gel (25 wt%) is 18.9 % (14.2/75). On the other hand, the PEG-water fraction of Tetra-PEG gel is 19.0 % (16/84).

Tensile tests and cyclic tests

Uniaxial tensile tests and loading-unloading tests of Tetra-PEG and SR gels were conducted on a Shimadzu AG-X plus universal tester (Shimadzu Corporation, Japan). At the strain rate of 0.125 s$^{-1}$, dumbbell-shaped specimens (JIS K 6251 No.8) were uniaxially stretched to rupture, or to certain extension ratio and then released to original length. The stress-extension ratio curves were automatically recorded by the universal tester.

Multiple cycles of loading-unloading tests were conducted on SR gels (25 and 50 wt%) to investigate their mechanical robustness. Stripe-shaped samples of 3 mm wide and 6 mm long were stretched and released within the range $1 < \lambda < 11$ for five cycles using Rheo Meter 150 ST (Sun Scientific Corporation, Japan). As shown in Fig. S1, the experiments were conducted in a special-ordered liquid paraffin chamber to prevent solvent evaporation.

Fracture tests

To estimate the fracture energy, pure-shear (PS) specimens of 35 mm wide, 5 mm long, and with a 15-mm initial crack introduced to the middle of short edge were uniaxially deformed in the height direction at the same strain rate with loading-unloading tests using the Shimadzu AG-X plus universal tester. Fracture processes were recorded by high-speed camera at 500 frames/sec to deduce crack velocity, and fracture energy $\Gamma$ was further calculated via equation (S1).

$$\Gamma = Wh \qquad (S1)$$

where $W$ and $h$ are respectively the strain energy density stored in un-notched sample right before fracture and the initial height of the sample. Fracture tests under similar condition were also conducted on single edge notched tension (SENT) specimens of 4 mm wide, 15 mm long, and with a 1 mm initial crack introduced to the middle of the long edge. A Nikon D3300 camera with macro lens was used to record the crack propagation in SENT specimen during the fracture test.

In-situ small angle/wide angle X-ray scattering tests

In-situ SAXS/WAXS were conducted at the beamline BL05XU of SPring-8, Japan, to characterize structural changes during the loading-unloading cycle. The wavelength of the X-ray was 1.0 Å and the sample-to-detector distances were 3950 mm and 300 mm, respectively for SAXS and WAXS tests. The samples were exposed to the X-ray for 1 s while being stretched by a homemade tensile machine. The stretching direction was horizontal, and the strain rate was 0.125 $s^{-1}$. The scattering patterns were obtained via 2D detector PILATUS 1M (DECTRIS, Switzerland). The transmittance was calculated from the ratio of the incident beam intensity measured by an ionization chamber to the transmitted beam intensity measured by a PIN diode embedded in the beam stopper.

The 2D WAXS patterns were converted to 1D intensity profiles from the direction perpendicular to stretching by symmetric-sector-averaging over azimuthal angles of 90° ± 2.5°. The averaged scattering intensity normalized by the transmission and sample thickness were plotted against the amplitude of the scattering vector $q$, defined as

$$q = \frac{4\pi}{\lambda} \sin\sin\left(\frac{\theta}{2}\right) \qquad (S2)$$

where $\theta$ is the scattering angle calibrated by the diffraction pattern of silver behenate (AgBe).

The 2D SAXS patterns were averaged by symmetric sector scanning and azimuthal scanning to plot the scattering intensity against the azimuthal angle and the scattering vector $q$, respectively, for further analysis based on the Ruland's method (23) and the Guinier's law (24).

A series of azimuthal scans of the perpendicular streaks were conducted at various q values, and the obtained profiles were fitted with a Lorentzian function. Representative results were given in Fig. S3(a). The average length $L$ and misorientation $B$ of SIC structure were described in the following equation:

$$B_{obs} = \frac{1}{L}\frac{1}{q} + B \qquad (S3)$$

where $B_{obs}$ was the integral width of the azimuthal peak of the streak and was obtained by dividing peak area by peak height. Thus, $L$ and $B$ could be respectively obtained from the slope and the intercept of the linear fitting of $B_{obs}$ - $1/q$ plot, as shown in Fig. S3(b). In this way, $L$ and $B$ of SIC structure were respectively estimated to be 100 nm and 2.7°, respectively.

Guinier law for rod-like scatter was applied to evaluate the radius $R$ of SIC structure. In Fig. S4, the logarithm of the 1D SAXS profile in the perpendicular direction $\ln[I(q)]$ was plotted against $q^2$. From the slope of linear portion at low $q$ region the cross-sectional gyration radius $R_c$ was estimated to be ~ 5.7 nm based on following equation:

$$\ln[I(q)] = \ln[I(0)] - \frac{R_c^2 q^2}{2} \qquad (S4)$$

Thus, the actual radius of the cylinder $R = \sqrt{2}R_c = 8$ nm.

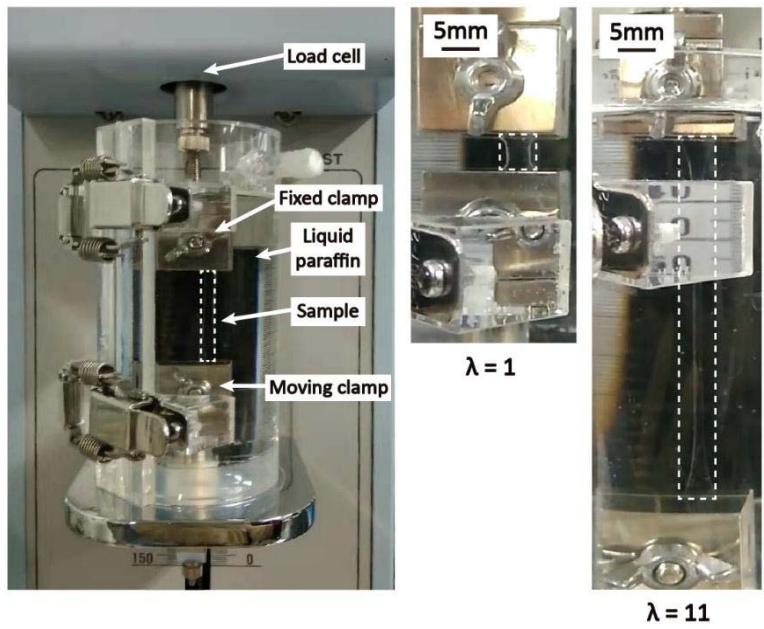

Fig. S1. Cyclic loading-unloading tests in the liquid paraffin chamber.

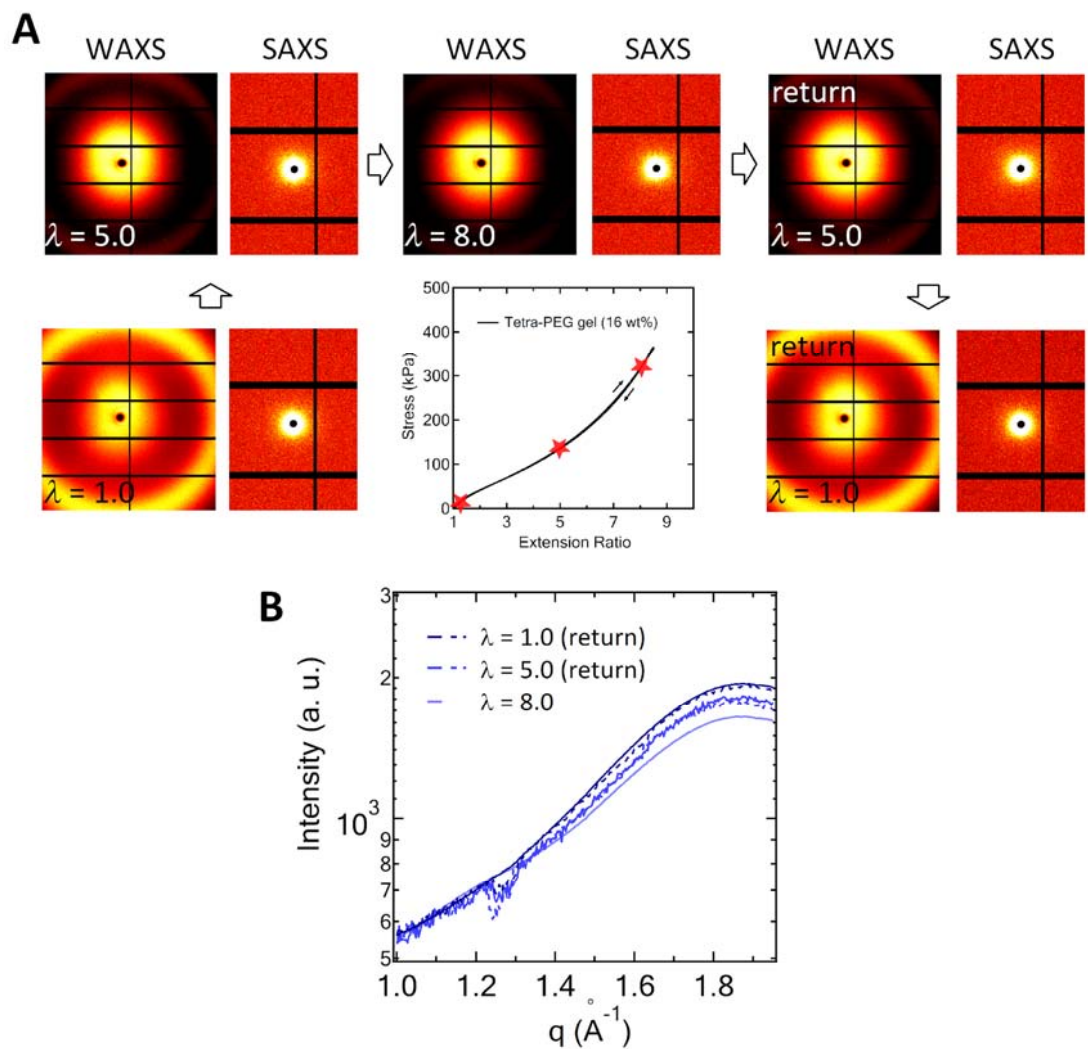

Fig. S2. Structural characterization of Tetra-PEG gel (16 wt%) under cyclic deformation. (A) 2D WAXS and SAXS patterns of Tetra-PEG gel obtained from the 1.0 < λ < 8.0 loading-unloading cycle. (B) 1D WAXS profiles of Tetra-PEG gel obtained via 5° symmetric sector average in the perpendicular direction.

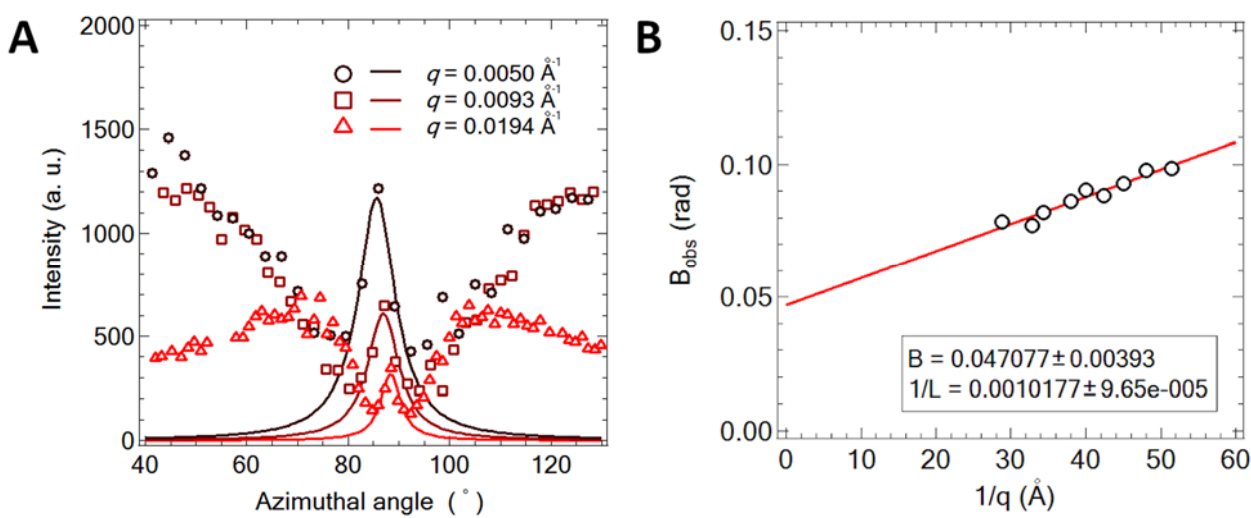

Fig. S3. Ruland's method utilized for estimating the average length $L$ and misorientation $B$ of SIC structure. (A) Azimuthal scanning and Lorentz fitting results of perpendicular streak at $q$ = 0.0050, 0.0093, and 0.0194 Å$^{-1}$; (B) Plot of integral width $B_{obs}$ vs $1/q$

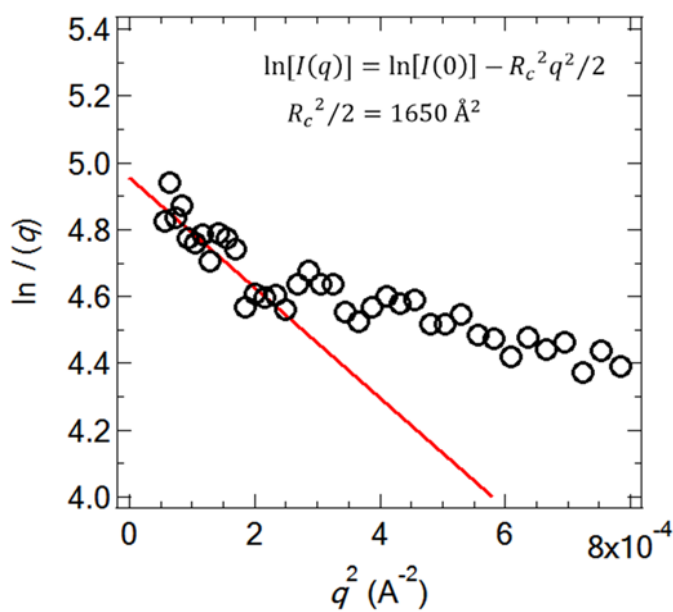

Fig. S4. Guinier law for rod-like scatter utilized for estimating the radius $R$ of SIC structure. ln $[I(q)]$ - $q^2$ plot of the perpendicular streak in the 2D SAXS pattern of SR gel (25 wt%).